\documentclass[usenatbib]{basi}
\usepackage[T1]{fontenc}
\usepackage[british]{babel}
\usepackage[varg]{txfonts}
\usepackage{natbib}
\usepackage[colorlinks,hypertexnames=false,citecolor=blue]{hyperref} 
\usepackage[figure,table]{hypcap} 

\begin{document}
\title[Magnetic structure of AR NOAA 11158]{Magnetic structure of solar active region NOAA 11158}
\author[P.~Vemareddy, A.~Ambastha \& T.~Wiegelmann]%
       {P.~Vemareddy$^1$\thanks{email: \texttt{vema@prl.res.in}},
       A.~Ambastha$^{1}$ and T.~Wiegelmann$^2$ \\
       $^1$Udaipur Solar Observatory, Physical Research Laboratory, Udaipur, India\\
       $^2$Max-Planck-Institut f$\ddot{u}$r Sonnensystemforschung, Katlenburg-Lindau, Germany}

\pubyear{2013}
\volume{41}
\pagerange{\pageref{firstpage}--\pageref{lastpage}}

\date{Received 2013 May 01; accepted 2013 September 01} 

\maketitle
\label{firstpage}

\begin{abstract}
Magnetic fields in the solar corona are responsible for a wide range of phenomena. However, any direct measurements of the coronal magnetic
fields are very difficult due to lack of suitable spectral lines, weak magnetic fields, and high temperatures. Therefore, one extrapolates photospheric field measurements into the corona. Owing to low coronal plasma $\beta$, we can apply a force-free model in lowest order to study the slow evolution of active region (AR) magnetic fields. On applying these models to AR 11158 and compared with coronal plasma tracers, we found that (1) the approximation of potential field to coronal structures over large length scales is a reasonable one, 2) linear force-free (LFF) assumption to AR coronal fields may not be applicable model as it assumes uniform twist over the entire AR, and 3) for modeling fields at sheared, stressed locations where energy release in the form of flares are usually observed, non-linear force free fields (NLFFF) seem to provide a good approximation. The maximum available free-energy profile shows step-wise decrease that is sufficient to power an M-class flare as observed.  
\end{abstract}


\section{Introduction} 
\label{s:intro}

Coronal loop structures, plasma eruption and acceleration of particles to high energies from the Sun provide evidences that the coronal structures and dynamics are controlled by solar magnetic fields. Sunspot or photospheric flux motions create currents and sites of reconnection in the chromosphere and corona. Active regions (ARs) of high magnetic complexity are also the sites of enhanced energetic activities such as flares and coronal mass ejections (CMEs). From a study of two recent ARs NOAA 11158 and 11166, \citet{vema2012a} showed that pumping of helicity by continuous flux motions leads to increasing complexity of magnetic structures. They inferred that monotonic injection of helicity accumulates in the corona that resulted in bursting of CMEs whereas injection of mixed signed helicity is prone to trigger flares. Further, \citet[][and references therein]{vema2012b} showed that the storage of magnetic energy, required for release in flares, occurs in the stressed magnetic field above the photosphere due to the photospheric and sub-photospheric motions. In other words, the energy to power the energetic transient activities is stored in the magnetic field structure of solar ARs, where the field must be at a higher energy state, or in non-potential state. Therefore, the extent to which the magnetic fields can be measured in successive layers above the photosphere directly and/or through comparison with modeling remains one of the key goals in solar physics. 

It is evident that the magnetic field in the solar corona is responsible for a wide range of phenomena. However, any direct measurements of the coronal magnetic fields are very difficult. The problems arise essentially due to (i) the lack of suitable spectral lines, (ii) weak magnetic fields, and (iii) large temperatures in the corona, which make measurements by Zeeman effect difficult. The coronal magnetic field strength is expected to be very small, around 10--50 G; therefore, the degree of circular polarization is expected to be very small. The low photon flux of the emission line solar corona ($\sim10^{-5}$ disk intensity) poses another major problem. The high coronal temperatures broaden the spectral lines, further reducing the Stokes V signal.  Therefore, measurement of the coronal Stokes V signal requires very sensitive polarimeter with extremely small instrumental polarization cross-talk.


Observations of the solar corona from space-borne experiments (e.g., TRACE, SDO) clearly demonstrate the magnetic fields through the ``tracers'' in the form of enhanced extreme ultraviolet (EUV) and X-ray emissions, which act as proxies of magnetic fields in the solar corona. However, EUV and soft X-ray images show only those magnetic field lines that carry significant density. They reveal how crucial the coronal magnetic fields are in almost all coronal activities.  In order to further advance our knowledge of the solar corona, however, direct measurement of the coronal magnetic field strength and configuration is indispensable. But, definitive quantitative measurement of the magnetic field is conspicuously missing.

In absence of reliable direct measurements of solar coronal magnetic fields, therefore, one needs to take the recourse to numerical modeling and/or or techniques of extrapolating the coronal fields from the available photospheric magnetic field observations.  The main aims of such studies are the following: (i) to reconstruct the coronal magnetic field that corresponds to an observed vector magnetic field measured in the photosphere, and (ii) to follow the slow evolution of a magnetic configuration through a series of states in response to footpoint motions. To achieve this goal, the logical first step is to seek a realistic representation of the solar magnetic field. That means, the configuration of the AR magnetic field needs to be modeled by using assumptions and approximations matching with the observed coronal features. Once we are able to construct the 3-D magnetic fields, one can study the storage and release of energy in the flares/CMEs. So, it is an important task to model the 3-D magnetic field of the AR and check if the model is a reasonable approximation of observed coronal magnetic field.

Extrapolation of photospheric magnetic fields into the corona can be carried out as a boundary value problem. For this purpose, we discuss in Section~\ref{s:meth} different models of extrapolations of coronal magnetic fields using the observed boundary conditions on the photosphere to model the 3-D structure of magnetic fields. Then, we apply these methods on the data set described in Section~\ref{s:data} to construct the magnetic structure of AR 11158 during transient flare period in Section~\ref{s:MagStr}. The resulting discussion and summary are given in Section~\ref{s:DisSum}. 

\section{Methods of magnetic field extrapolations}
\label{s:meth}
The equilibrium structure of a system of magnetic field and plasma on the Sun can be described by the equation of magneto hydrostatics,
\begin{equation}-\nabla p+\frac{1}{4\pi }(\nabla \times \mathbf{B})\times \mathbf{B}-\rho \nabla \Phi =0
\end{equation}
Here p is the plasma pressure, \textbf{B} is the magnetic field, $\rho$ is the density, and $\Phi$ is the gravitational potential on the Sun's atmosphere. In the chromosphere and corona above the ARs, the effect of the magnetic field generally overwhelms the pressure and gravity forces (plasma $\beta<<1$) so that the equilibrium is approximately
\begin{equation}
(\nabla \times \mathbf{B})\times \mathbf{B}=0;
\end{equation}
This equation defines the force-free magnetic field, in which Lorentz force balances by itself. From Ampere's law, $\nabla \times \mathbf{B}=\mu \mathbf{J}$ where \textbf{J} is current density, the force free equation can also be written in terms of currents $\mathbf{J}\times \mathbf{B}=0$. The force-free equation is a non-linear one in \textbf{B} and difficult to find any satisfactory analytical solutions. The simple solution to satisfy this equation is to assume that the current \textbf{J} is parallel to \textbf{B}, i.e,
\begin{equation}
\nabla \times \mathbf{B}=\alpha \mathbf{B};\,\,\,\,\,\nabla \bullet \mathbf{B}=0
\end{equation}
with the force-free function $\alpha$. Therefore, the computation of force-free fields subdivides the problem into the mathematically simpler cases: 1) Potential fields with $\alpha = 0$ resulting in divergence free condition to Laplace equation, 2) Linear force-free fields (LFF) $\alpha = {\rm constant}$ , and 3) Non-Linear Force-Free fields (NLFFF) in which $\alpha$ is a function of space.

Modeling solar coronal magnetic field with these equations is a boundary value problem. As lower boundary in our computational box we use the measured photospheric
magnetic field of flaring AR. Given normal or line-of-sight (los) component magnetic field, coronal potential fields can be constructed at each grid point over the AR by solving the Laplace equation \citep{schmidt1964}. For linear force-free fields, twist or torsion parameter $\alpha$ is needed in addition to normal field \citep{gary1989}.

Non-linear force-free fields are more general and require all three components of magnetic field observations on the boundary as each field line has its own twist i.e., $\mathbf{B}\bullet \nabla \alpha =0$. Since we need Lorentz force of magnetic field to be zero, minimizing the functional having force-free and divergence free conditions  iteratively with boundary conditions \citep{wheatland2000,wiegelmann2004}, will lead to the required NLFFF fields over the AR. To mitigate the effect of errors and lack of data on the constructed field, the explicit functional minimization \citep{wiegelmann2010,wiegelmann2012} involves:
\begin{equation}
L=\int\limits_{V}{\left( w\frac{|(\nabla \times \mathbf{B})\times \mathbf{B}{{|}^{2}}}{{{B}^{2}}}+w|\nabla \bullet \mathbf{B}{{|}^{2}} \right)}dV+\nu \int\limits_{S}{\left( \mathbf{B}-{{\mathbf{B}}_{obs}} \right)}\bullet W\bullet \left( \mathbf{B}-{{\mathbf{B}}_{obs}} \right)dS
\label{e_Opt_sl_inj}
\end{equation}
The first integral contains force-free and divergence conditions in quadratic forms and are obviously fulfilled when the functional reaches its minimum at $L = 0$ where the boundary conditions entered directly as iterative improvements of \textbf{B} to minimize $L$ were constrained to $B = B_{\rm obs}$ at the photospheric boundary. The surface integral over the photosphere is to take into account the errors and lack of data on boundary while iterating \textbf{B} by injecting boundary observations with the speed indicated by $\nu$ to minimize $L$. The weight function W(x, y) is a diagonal error matrix, the elements $W_{\rm los}$, $W_{\rm trans}$, $W_{\rm trans}$ of which are chosen inversely proportional to the local measurement error of the respective photospheric field component at x, y.

Even if we choose a sufficiently flux-balanced isolated AR the force-free conditions \citep{aly1989} are not fulfilled for measured vector magnetograms consistent with NLFFF algorithm. So, preprocessing the photospheric vector data to drive towards near force-free conditions using the freedom of large noise in transverse components was proposed \citep{wiegelmann2006}.

\section{The observational data}
\label{s:data}
The {\it Helioseismic and Magnetic Imager} (HMI; \citet{schou2012}) on board the {\it Solar Dynamics Observatory} (SDO) gives us an unprecedented opportunity to better understand the flares' energetics using its high-quality photospheric vector magnetograms (VMGs).

For the required analysis of magnetic structure of AR NOAA 11158, we obtained HMI VMGs of 12 minutes cadence spanning 4 h around the M6.6 flare at 17:12UT on 13 February 2011. The HMI instrument observes the full Sun at six wavelengths in the Fe {\sc i} 6173 \AA~photospheric absorption line. Filtergrams at 0.5 arcsec/pixel ($\sim362$~km) resolution are processed to obtain Dopplergrams, line of sight(LOS), and vector magnetograms. Stokes parameters derived from filtergrams taken at 135 s cadence are averaged to 12 minutes. These are then inverted to derive vector magnetograms by Milne-Eddington inversion algorithm \citep{borrero2011}. The  $180^{\rm o}$ azimuthal ambiguity in the transverse component is resolved using an improved version of the minimum energy algorithm \citep{metcalf1994, leka2009}. After correcting for projection effects \citep{venkat1989,gary1990} the data were re-mapped and transformed to heliographic coordinates from spherical coordinates using a Lambert equal-area projection \citep{calabretta2002}. For a detailed description of all this data processing from level-0 filtergrams to level-3 vector magnetograms, please refer to \citet{hoeksema2012}.

In order to examine how well the extrapolated field represents the real coronal field, we compared the modeled field lines with coronal plasma loops as they are ultimate tracers of particles along underlying magnetic field lines. For this purpose, we used EUV imaging data from Atmospheric Imaging Assembly \citep{lemen2012} on board SDO, having 0.6"/pixel resolution and 12 s cadence.
\begin{figure}[ht!]
\centering
\includegraphics[width=.8\textwidth,clip=]{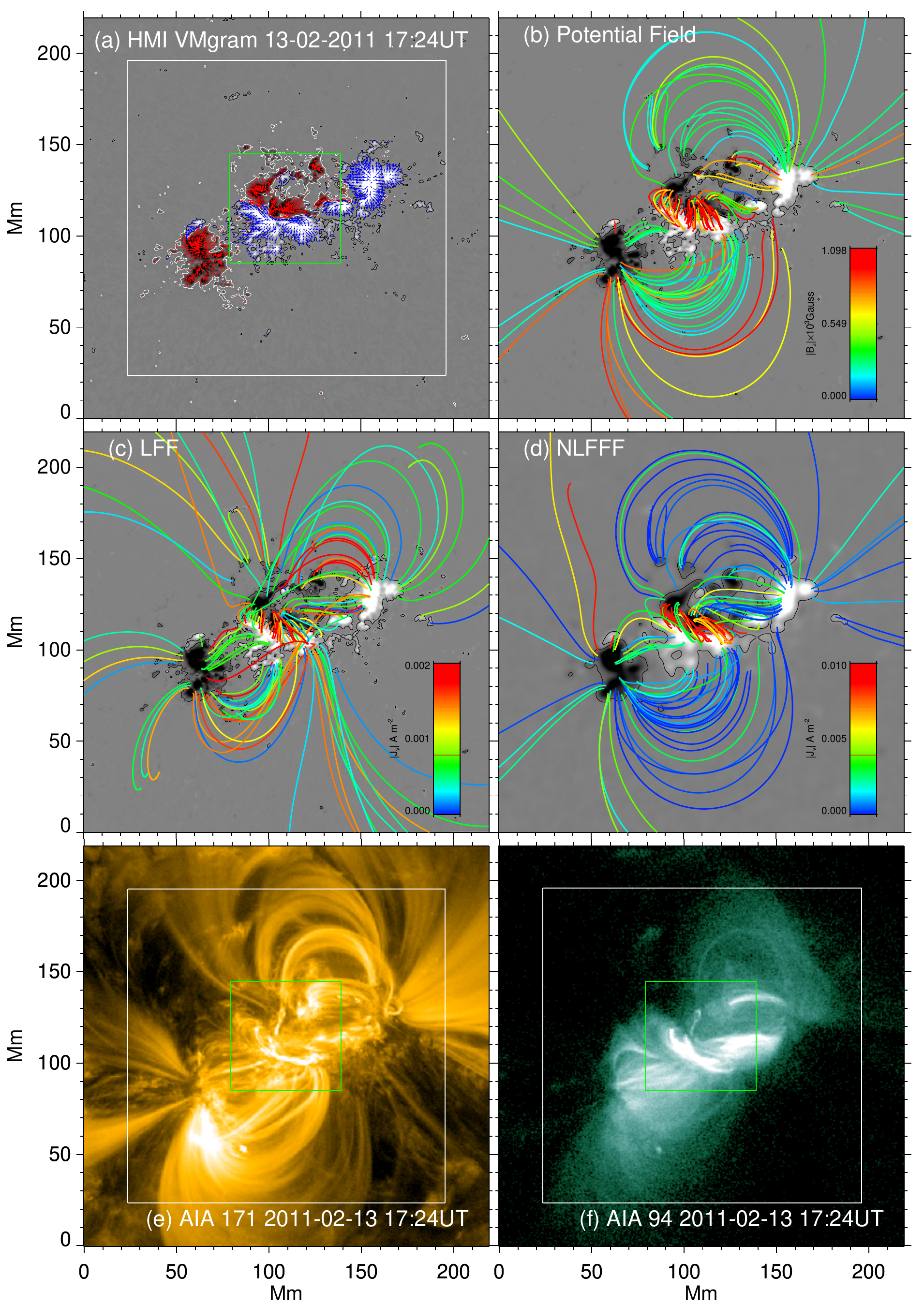}
\caption{Magnetic field structure of AR 11158 with projected field lines onto 2-D
view using different models of extrapolations. The field lines are colored according to
the scale of vertical current (vertical field) in panels (c)-(d) (panel (b)), respectively. Coronal observations in 171 and 94~\AA~channels as proxies of coronal magnetic fields are shown in (e)-(f) panels }\label{Fig_11158}
\end{figure}

\section{Magnetic structure of AR 11158} 
\label{s:MagStr}
The AR NOAA 11158 was a newly emerging region that first appeared as small pores on the solar disk on February 11, 2011 at the heliographic location E33$^{\rm o}$S19$^{\rm o}$. Subsequently it grew rapidly by merging of small pores and formed bigger sunspots developing into magnetic complexity on 2011 February 13. The HMI intensity maps show the spatio-temporal evolution with large proper motion of individual sunspots during February 13-16. From an inspection by overlying LOS contours on intensity maps, it was clear that the overall magnetic configuration of AR as quadrupolar nature. This AR and its flare/CME productivity have been thoroughly investigated \citep{xudong2012,vema2012a,vema2012b}. 

Fig.~\ref{Fig_11158}(a) portrays VMG of this AR at 17:24UT with transverse $B_t$ vectors on $B_z$ image within contour level of 150 (-150) G indicated by green (red) arrows.  Although, the spatial dimensions of the VMGs enclosing the AR are isolated from surrounding flux (${{\in }_{flux}}=0.02$), the force-free conditions (${{\in }_{force}}=0.057<<1;{{\in }_{torque}}=0.05$) are not well satisfied. Therefore, these bottom boundaries of VMGs are preprocessed by assigning a uniform altitude of 720 km (1 pixel) to this layer to suit for force-free extrapolation. To reduce the computational difficulties and noise, the AR was rebinned to $300\times300$ pixel$^2$ and computational grid of $300\times300\times160$ was used to represent the physical solar dimensions of $219\times219\times160$~Mm$^3$ encompassing the AR.

We first calculated potential field on the above computational grid by setting $\alpha=0$ \citep{gary1989} with $B_z$ component of $\textbf{B}_{\rm obs}$ as boundary condition. The potential field lines at selected locations are integrated and plotted (Fig.~\ref{Fig_11158}(b)) over the lower bounary of $B_z$. For better visualization, we colored the field lines according to the magnitude of LOS field at their footpoint positions. 

In order to know the suitability of a particular model, we also constructed the 3-D field under the LFF assumption. In this assumption, whole AR is twisted by certain constant amount characterized by the twist parameter $\alpha$. We estimated this twist parameter as $\alpha_{\rm best}=(0.0137\pm0.0014)\times10^{-6}$~m$^{-1}$ by an iterative algorithm described in \citet{vema2012b}. Essentially, for a given $\alpha$  and $B_z$ field, we search for $\alpha_{\rm best}$ at which the force-free extrapolated transverse field ($B_t$) best fits the observed transverse field in a sense of minimum least squared difference on the photosphere. With this best $\alpha$, we constructed linear force-free field at each grid point in 3-D box and plotted the field lines at the same locations as the potential approximation in Fig.~\ref{Fig_11158}(c).

Then, the calculated potential field is given as initial field to NLFFF algorithm \citep{wiegelmann2012} with the observed 3 preprocessed components of field $\textbf{B}_{\rm obs}$ as lower boundary condition by setting 32 pixel widths as buffer boundary of cosine profile to reduce the effect of unknown lateral and top boundaries on the evolved field. A weight function W(x,y) (second term in Equation~\ref{e_Opt_sl_inj}) proportional to the square of the magnitude of horizontal field is used with the slow injection of boundaries at a rate set by $\nu=0.005$. While iteration, the following quantities are monitored.
\begin{equation}
L_1=\int\limits_{V}{\frac{|(\nabla \times \mathbf{B})\times
\mathbf{B}{{|}^2}}{{{\rm B}}^2}}dV;\,\,L_2=\int\limits_{V}{|(\nabla
\bullet \mathbf{B}{{|}^{2}}}dV;\,\,\sigma_j=\left(
\sum\limits_{i}{\frac{|{\rm J}\times
{{\rm B}_{i}}|}{{\rm B}_{i}}}
\right)/\sum\limits_{i}{\rm J}_{i}
\end{equation}
where $L_1$ and $L_2$ correspond to the first and second terms in Equation~\ref{e_Opt_sl_inj}, respectively, in the inner $236\times236\times128$ physical region where weighting functions are unity and sine of the current weighted average angle   between the magnetic field and electric current. Success of the algorithm is ensured by progressive decrease of these terms forcing current parallel to magnetic field (force-free) quantified by $\sigma_{j}$ reaching to around $10^{\rm o}$ in our calculations on the data set. The corresponding NLFFF field lines are plotted over the $B_z$ map in Fig.~\ref{Fig_11158}(d). The SDO-AIA observations in EVU channels 171, 94Å which are proxies or plasma tracers of coronal magnetic fields of the AR are plotted in panels (e)--(f) in the same figure.

Long overlying loops in the coronal observations are seen to be explained best with the potential and NLFF models, but the LFF model does seem to give a worst approximation. We can distinguish the difference between the NLFFF and potential field structure in the central part of the AR within the green square box, where most sheared or twisted structures are present. \citet{vema2012a,vema2012b} found large shear motion of fluxes about the polarity inversion line by which they speculated the reason for repeated flare activity at this location. 

\begin{figure}[ht!]
\centering
\includegraphics[width=.8\textwidth,clip=]{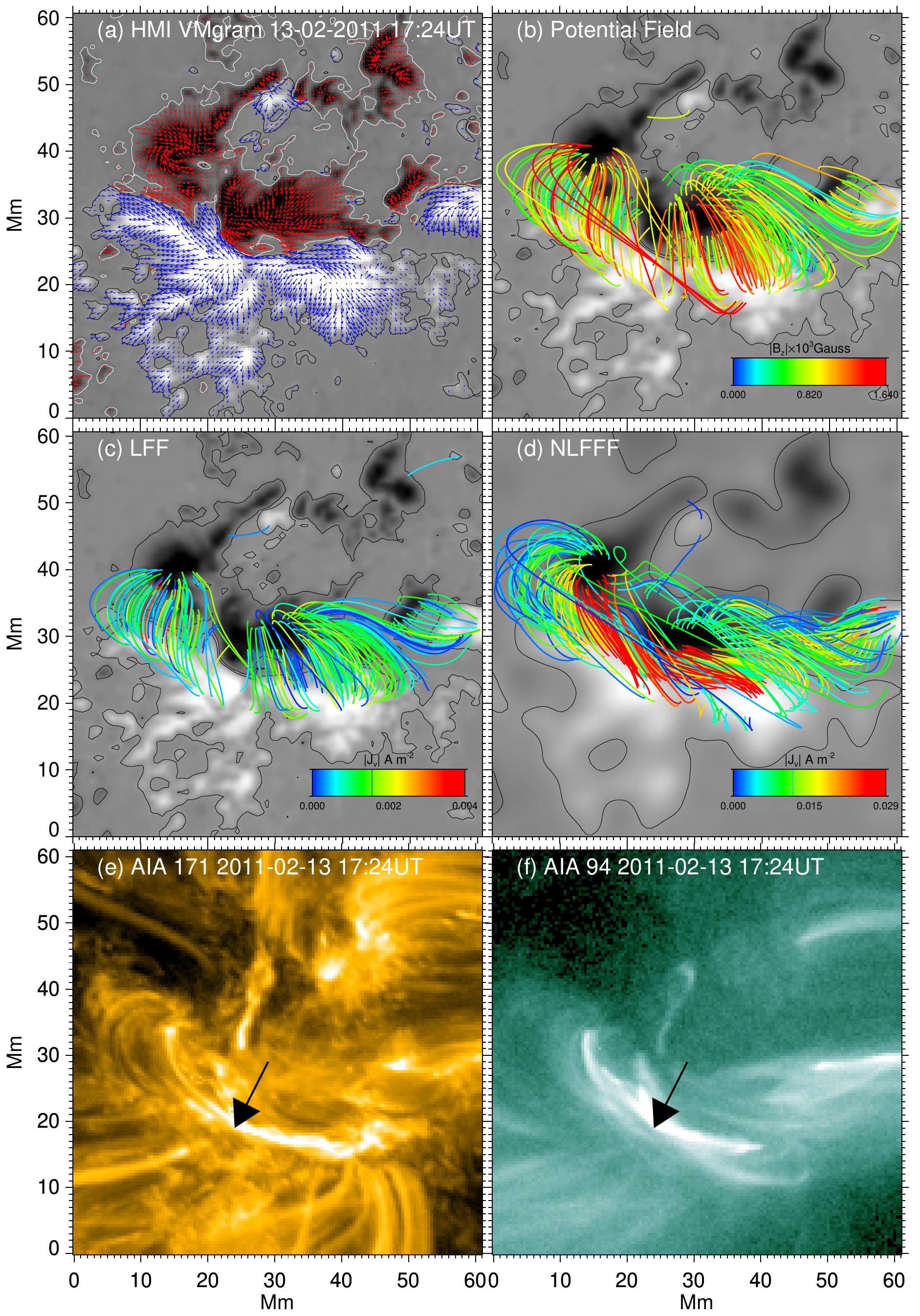}
\caption[Twisted flux system in AR 11158]{The central portion [region of green box in Fig.~\ref{Fig_11158}(a)] of AR 11158 which consisted of highly twisted flux system, appearing as bright plasma emission in 94 \AA. In all 3 models, the same footpoints of field lines are chosen, but only closed field lines are kept.}\label{Fig_11158_roi}
\end{figure}

In Fig.~\ref{Fig_11158_roi}, we have focused further on this twisted central region. The field lines are selected at pixels above 800G of transverse ($|B_t|$) field at one foot point and only those landing on the surface (closed field lines) are kept. The same field lines are drawn using the potential field, LFF and NLFFF models in frames (b) to (d)). For better visualization, we color scaled the field lines according to LOS field (b) and vertical current (c-d) at their foot points. The blowups of SDO-AIA observations in EVU channels 171, and 94 \AA~corresponding to the green box in Fig.~\ref{Fig_11158} are also shown in Fig.~\ref{Fig_11158_roi}(e-f).

The alignment of transverse vectors with the polarity inversion line (PIL) delineates presence of strong shear that occurred due to continuous footpoint shear motion. The consequent stressing of the field lines stored magnetic energy in the system. In the potential field case, they simply connect pairs of opposite polarities orthogonally to the PIL. In LFF, the entire region (of one sign) is assigned by particular constant $\alpha$, which is apparently not the case in complex ARs. Generally, each field line is expected to have its own twist due to the current flowing along it, and therefore must be specified by different $\alpha$, rather than a constant $\alpha$. Such a provision is made with NLFFF algorithm, as field lines having different currents appear as twisted flux bundle above the interface of positive and negative flux regions. The magnetic structure of NLFFF field shows twisted flux system connecting foot points on either side of PIL under the potential (less current flowing field lines) field arcade. Such a system represents stored configuration of energy in the form of twist indicated by current flowing along the underlying field lines. Bright plasma emission in EVU channels (pointed by black arrow) is observed surrounding this twisted flux system of current. Therefore, highly complex, twisted flux regions are the locations of stored free-energy required to power flare, and are better reproduced by NLFFF models than by potential and LFF models.  

\begin{figure}[ht!]
\centering
\includegraphics[width=.9\textwidth,clip=]{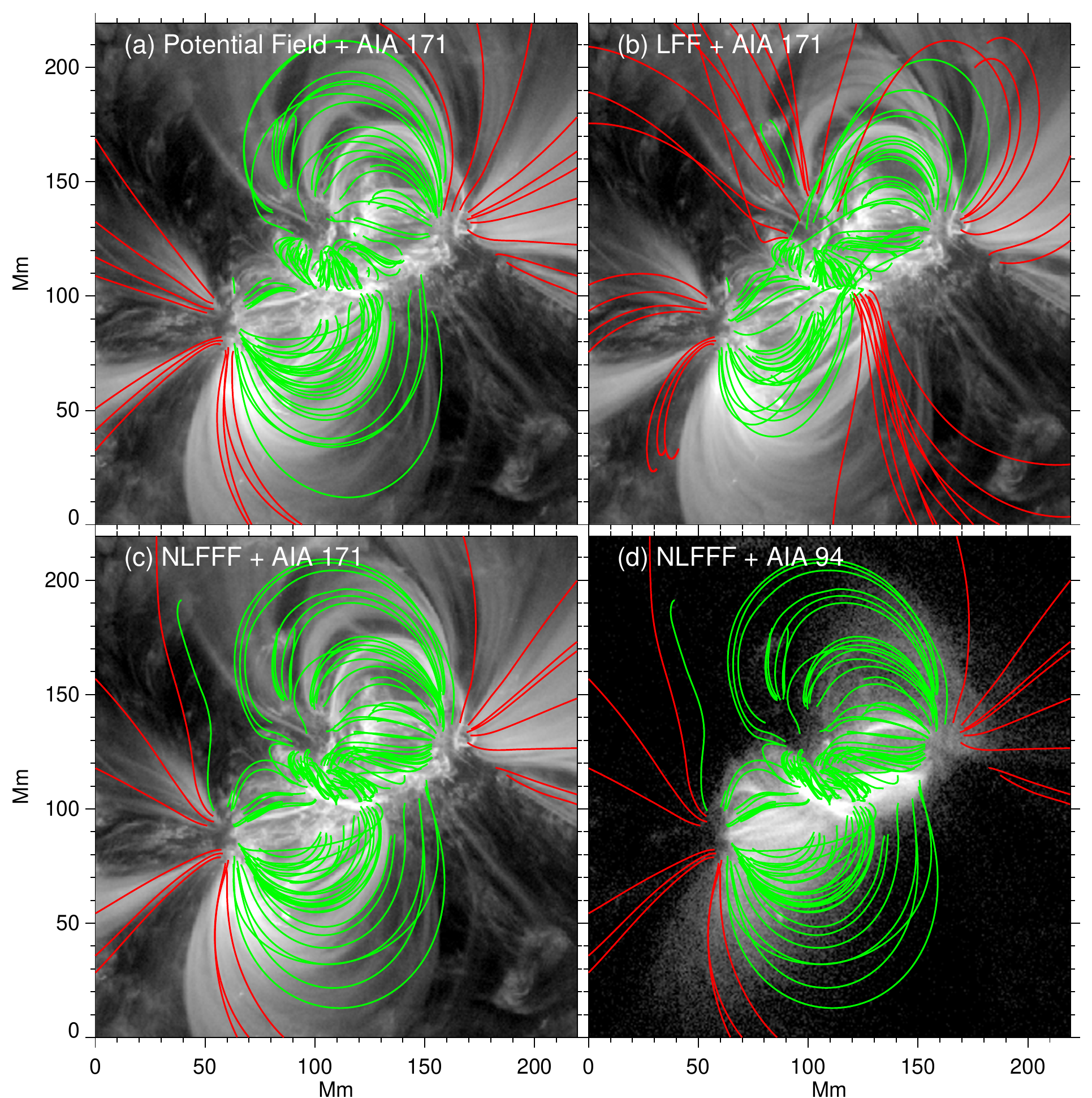}
\caption[Comparison of Models]{Comparison of extrapolation models to the observed plasma tracers of coronal magnetic field. In panels (a)-(c), the background image is EVU observation in 171~\AA~and in (d) is soft X-ray 94\AA~from AIA. Closed (open) field lines are drawn with green (red) color.}
\label{Fig_11158_compare}
\end{figure}

In Fig.~\ref{Fig_11158_compare}, we have compared the potential, LFF and NLFFF modeled field lines with the plasma tracers by overlying them on EUV 171, and 94 \AA~images. Many field lines in potential and NLFFF models best resemble the plasma tracers in panels ((a), (c), (d)) except at the central portion where shearing flux motions dominated. However, there is no such similarity seen between LFF field lines and plasma tracers from panel (b). Reproducing the fields at locations having shear motions, would be the most challenging task for the reasons of estimating free-magnetic energy from that location. To some extent, we see that the NLFFF model generated structures similar to the coronal observations (tracers), and would be a promising coronal magnetic field model.

The magnetic free-energy is the maximum energy available to the system to power flares above potential energy ($E_{\rm P}$) state of that system. This can be estimated by subtracting potential field energy from NLFFF energy ($E_{\rm NLFFF}$):
\begin{equation}
E_{\rm free}=\int\limits_{V}{\frac{\mathbf{B}_{\rm NLFFF}^{2}}{8\pi }}dV-\int\limits_{V}{\frac{\mathbf{B}_{\rm P}^{2}}{8\pi }}dV
\end{equation}
Using this equation, we calculated magnetic free energy of the AR before (at 17:12UT) and after (at 17:48UT) the M6.6 flare. We note that the ratio of $E_{\rm NLFFF}/E_{\rm p}$ decreased from 1.13 before flare to 1.10 after the flare in a step-wise manner. Therefore, the free-energy released during this M6.6 flare is calculated to be $0.97\times10^{31}$ ergs in the computational domain above the AR. Further, the same free-energy estimated from Virial theorem encompassing only twisted flux (green box) gives a higher value $4.3\times10^{31}$ ergs \citep{vema2012b}. In both cases, there are possibilities of under/over estimation of the energy because of finite divergence of field lines from chosen area of AR.

\section{Discussion and summary}
\label{s:DisSum}
Using the available models and procedures in the literature, we have described the working concepts of extrapolations of coronal magnetic fields from the observed boundary conditions. The main approximation involved is magnetic field dominated solar atmosphere over plasma pressure and gravity, i.e., the so-called force-free assumption.

Specifying the normal component of boundary condition from observations, the three components of magnetic fields can be constructed at each grid point over the AR under study for obtaining magnetic structure by the assumption of potential and linear force-free approximations. However, these assumptions are usually beyond the real situations of solar magnetic fields, and may not be suitable for explaining the observed coronal plasma structures. Therefore, many nonlinear force-free algorithms have been proposed and applied to solar ARs. These methods require high quality photospheric VMGs for use as boundary conditions, such as, those available recently from SDO-HMI. Using the optimization algorithm, we have attempted to explain magnetic structure of an AR observed by HMI. From this study of AR 11158, we note the following major observations:

\begin{enumerate}

\item We infer that the approximation of potential field to coronal structures over large length scales is a reasonable one but it is not at all a useful model for compact twisted regions. LFF assumption to AR coronal fields may not be applicable model as it assumes uniform twist over the entire AR.

\item For modeling fields at sheared, stressed locations where energy release in the form of flares are usually observed, NLFFF models seem to provide a good approximation.

\item AR NOAA 11158 possessed a complex $\delta$ magnetic configuration, having stressed fields by shearing flux motions about the PIL where a large M6.6 flare occurred. The NLFFF field lines computed for this AR reveal presence of inner twisted flux system giving rise to currents where the bright plasma emission in EUV is seen.

\item Estimation of magnetic free energy i.e., the energy above potential field energy, shows step-wise decrease during the M6.6 flare releasing $0.97\times10^{31}$ ergs sufficient to power an M-class flare.
\end{enumerate}

There are points to be noted that may lead to inappropriate results as a consequence of a bad model. The preprocessing scheme oversmooths the currents and hence lowers the twist of field lines and structures may not resemble well with coronal observations. All of these models assume at or planar boundary of observations which is not the case with the real Sun over a large field of view. For example, a magnetogram spanning about $18^{\rm o}$ in Heliographic longitude, as used in this study, will then have its edge elevated by $(1-\cos 9^{\rm o}$)$R_\odot$($\sim8.7$Mm or 12 pixels) above the solar surface. This may also give rise to incorrect estimates of fields in reproducing the loop structures.

In applying the extrapolation procedures, one should keep the above cautions while judging how well the model suits the observed structures. The NLFFF models appear to work reasonably well in reproducing twisted loops. Therefore, we plan to use them in our future work for the study of evolution of magnetic structure and explosive phenomena in solar ARs.

\section*{Acknowledgements}
The authors thank the science teams of NASA/SDO for their efforts in obtaining the data. This work uses the high performance computing facility at PRL computer center and we thank Mr. Jigar Raval for his technical help in this regard.


\label{lastpage}
\end{document}